# Customising radiative decay dynamics of two-dimensional excitons via position- and polarisation-dependent vacuum-field interference

Sanghyeok Park[1,^], Dongha Kim[1,^], Yun-Seok Choi[2], Arthur Baucour[3], Donghyeong Kim[1], Sangho Yoon[4,5], Kenji Watanabe[6], Takashi Taniguchi[7], Jonghwa Shin[3], Jonghwan Kim[4,5], and Min-Kyo Seo[1]*

^: equally contributed

*minkyo_seo@kaist.ac.kr

1. Department of Physics, KAIST, Daejeon, Daehak-ro, 291, 34141, Republic of Korea
2. Department of Chemistry, KAIST, Daejeon, Daehak-ro, 291, 34141, Republic of Korea
3. Department of Materials Science and Engineering, KAIST, Daejeon, Daehak-ro, 291, 34141, Republic of Korea
4. Department of Materials Science and Engineering, Pohang University of Science and Technology, Pohang, Republic of Korea
5. Center for van der Waals Quantum Solids, Institute for Basic Science (IBS), Pohang, Republic of Korea
6. Research Center for Functional Materials, National Institute for Materials Science, Tsukuba, Ibaraki, Japan
7. International Center for Materials Nanoarchitectonics, National Institute for Materials Science, Tsukuba, Ibaraki, Japan

**Introductory Paragraph**

Embodying bosonic and electrically interactive characteristics in two-dimensional space, excitons in transition-metal dichalcogenides (TMDCs) have garnered considerable attention[1]. The realisation and application of strong-correlation effects[2,3,4,5,6,7,8], long-range transport[9,10,11,12], and valley-dependent optoelectronic properties[13,14,15,16] require customising exciton decay dynamics. Strains[17,18], defects[19], and electrostatic doping[20,21,22] effectively control the decay dynamics but significantly disturb the intrinsic properties of TMDCs, such as electron band structure and exciton binding energy. Meanwhile, vacuum-field manipulation provides an optical alternative for engineering radiative decay dynamics. Planar mirrors[23,24,25,26] and cavities[27,28,29] have been employed to manage the light–matter interactions of two-dimensional excitons. However, the conventional flat platforms cannot customise the radiative decay landscape in the horizontal TMDC plane or independently control vacuum field interference at different pumping and emission frequencies. Here, we present a meta-mirror resolving the issues with more optical freedom. For neutral excitons of the monolayer MoSe$_2$, the meta-mirror manipulated the radiative decay rate by two orders of magnitude, depending on its geometry. Moreover, we experimentally identified the correlation between emission intensity and spectral linewidth. The anisotropic meta-mirror demonstrated polarisation-dependent radiative decay control. We expect that the meta-mirror platform will be promising to tailor the two-dimensional distributions of lifetime, density, and diffusion of TMDC excitons in advanced opto-excitonic applications.

The meta-mirror consists of an Au bottom mirror, an SiO$_2$ spacer, an Au nanodisk array, a spin-coated hydrogen silsesquioxane (HSQ) superstrate, and a hexagonal Boron Nitride (hBN) layer (Fig. 1a). TMDC materials are located on top of the hBN layer of the meta-mirror and might be covered by another thin hBN layer (Methods). Depending on the radius and period of the Au nanodisks, the plasmonic meta-mirror reflects the incident light with different phases at the pumping and exciton emission frequencies[30,31,32,33] (Supplementary Information S1). We optimised the thickness of the SiO$_2$ spacer to determine the combination of the radius and period of the Au nanodisks, which supports an identical reflection amplitude at a given frequency of light. The reflection phase ($\phi_r$) and amplitude ($|r|$) of the light determine the vacuum field interference in the plane of interest along the vertical direction, including the top surface of the meta-mirror. The stronger the vacuum field, the faster will be the radiative decay, and the higher will be the density of excitons. By arranging different meta-mirrors, we can yield two-dimensional lithography of the exciton density. A schematic illustration of two representative meta-mirrors (MM1 and MM2) that operate identically for pumping but distinctly for emission is depicted in Fig. 1b. At the pumping frequency, MM1 and MM2 possess the same reflection amplitude and phase ($\phi_{r1} = \phi_{r2}$) and support identical interference between the incident and reflected light. At the emission frequency, MM1 and MM2 reflect the incident light with the same amplitude but different phases ($\phi_{r1} \neq \phi_{r2}$). Depending on the reflection phase, the vacuum-field interference at the PL emission frequency can be controlled between constructive and destructive conditions.

For the experiment, we prepared three different plasmonic meta-mirrors (A, B, and C) to demonstrate optical customisation of the radiative decay dynamics of excitons (Fig. 1c and Supplementary Information S2). The spin-coated HSQ superstrate gets rid of the surface fluctuation from the Au nanodisk array and minimised the local strain effects on the employed two-dimensional material (Supplementary Information S3). We targeted the neutral excitons of the monolayer MoSe$_2$, of which the photoluminescence (PL) emission frequency was 1.63 eV, and the pumping frequency was 2.33 eV. The reflection amplitude and phase of the meta-mirrors were examined by off-axis holography (Fig. 1d and Methods). Here, we measured the reflection phase shift ($\Delta\phi_r$) of the meta-mirror relative to that of the region without Au nanodisks. At the pumping frequency, the three meta-mirrors demonstrated approximately the same reflection amplitude and reflection phase, ensuring identical exciton excitations. However, at the PL emission frequency, the reflection phase exhibits a stepwise change among the different meta-mirrors while maintaining the reflection amplitude proximate to unity. Notably, the higher the reflection amplitude, the higher will be the visibility of the vacuum field interference, and the higher will be the dynamic range in the radiative decay engineering (Supplementary Information S4). The reflection amplitude and relative phase averaged over the area of each meta-mirror are plotted in Fig. 1e. The relative reflection phase at the PL emission frequency changes by 69.47°, from 29.49° to 98.96°. A phase change of π is required for covering all ranges between the constructive and destructive interference of the vacuum field. The numerical calculations of the meta-mirror indicate that a maximum relative phase change of 269.59° can be achieved, depending on the radius and period of the Au nanodisks (Supplementary Information S2).

The meta-mirror enables robust control of the radiative recombination of excitons and the two-dimensional arrangement of areas featuring different properties of radiative decay. By employing a confocal microscopy setup (Methods), we initially measured the PL intensity ($I_{PL}$) distribution of the $MoSe_2$ monolayer's neutral excitons on the Si/SiO$_2$ substrate and the three neighbouring meta-mirrors presented in Figs. 1c–1e. We prepared the $MoSe_2$ monolayer flake encapsulated by the hBN layers with thicknesses of 6 and 174 nm, of which the thicker layer is the top layer of the meta-mirror platform. The $MoSe_2$ monolayer, placed on an Si/SiO$_2$ substrate, exhibits uniform PL over its entire area (Fig. 2a). However, after transfer onto the meta-mirrors, the PL intensity varies significantly and sharply for each meta-mirror (Fig. 2b). The ratio of the averaged PL intensities on the three different meta-mirrors (A, B, and C) is 9.78:4.70:1.00. Considering that the employed meta-mirrors support pumping conditions that are approximately identical (Figs. 1d and 1e), the manipulation of the vacuum field interference at the emission frequency dominates the difference in the PL intensity. The three meta-mirrors enabled the control of the vacuum field and exciton PL emission intensity by one order of magnitude. In addition, the arrangement of different meta-mirrors provides a platform to spatially engineer light-matter interaction properties in two dimensions as desired, beyond the limits of typical planar mirrors.

To further reveal changes in the radiative decay dynamics of neutral excitons via vacuum field interference manipulation, we statistically analysed the intensity and linewidth of the PL spectrum of the $MoSe_2$ monolayer on the SiO$_2$/Si substrate (white dashed rectangle in Fig. 2a) and the meta-mirrors (white solid rectangles in Fig. 2b). Given the PL spectrum with a Lorentzian shape, the precisely measured linewidth corresponds to the total decay rate ($\gamma_{tot}$) of excitons. As plotted in Fig. 2c, the pristine $MoSe_2$ monolayer on the SiO$_2$/Si substrate exhibits an ordinary statistical distribution in which the linewidth is 1.58±0.21 meV. The centre frequency of the neutral exciton emission of the pristine $MoSe_2$ monolayer is 1.645 eV (the inset of Fig. 2c). However, the statistics of the PL spectra on the meta-mirrors indicate a high correlation between the PL intensity and linewidth, as demonstrated in Fig. 2d. The meta-mirrors A, B, and C result in distinct linewidths of 2.24, 1.53, and 1.28 meV in median, respectively, which reflect the change in the exciton lifetime owing to the manipulation of the vacuum-field interference. Moreover, the standard deviations of the PL linewidths measured on the meta-mirror A, B, and C, are 0.25, 0.22, and 0.27 meV, respectively, and their values normalised to the averaged linewidths are 10.88%, 13.82%, and 19.61%. Considering that the normalised standard deviation of the PL linewidth of the pristine $MoSe_2$ monolayer on the $SiO_2$/Si substrate was 13.15%, the meta-mirrors seamlessly preserved the inherent properties of the two-dimensional material without causing undesired local strains or deformations. Fig. 2e shows the representative PL spectra of neutral excitons engineered by the meta-mirror A, B, and C. We expect that the slight emission frequency change of ~3.7 meV originated from the cooperative Lamb shift[24] owing to the coherent interaction between the vacuum-field fluctuations and excitons (Supplementary Information S5).

The measured PL statistics can be theoretically modelled by considering coherent optical feedback[23,24,25,26]. The reflection amplitude and phase pickup engineered by the meta-mirror control the local density of optical states and the Purcell factor ($F_P$) at the $MoSe_2$ monolayer. Consequently, the total decay rate of excitons is expressed as $\gamma_{tot} = \gamma_{nrad} + F_P\gamma_{rad,0} = \gamma_{tot,0}[1 + \eta_{int}|r|\cos(\phi_r)]$, where $\eta_{int}$ represent the internal quantum efficiency of radiative recombination (Supplementary Information S4). The internal quantum efficiency relates the intrinsic total and radiative decay rates, $\gamma_{tot,0}$ and $\gamma_{rad,0}$, respectively, of the monolayer $MoSe_2$ as $\gamma_{rad,0} = \eta_{int}\gamma_{tot,0}$. Fig. 2f depicts the theoretical and experimental results for the total decay rate as a function of the measured relative reflection phase. The model analysis of the measured total decay rates extracted the internal quantum efficiency and non-radiative decay rate of $MoSe_2$ monolayer's excitons to be 0.51 and 1.27 meV in median, respectively. The extracted non-radiative decay rate explains the correlation between the PL intensity and linewidth following $I_{PL} \propto \gamma_{rad}/\gamma_{tot} = (\gamma_{tot} - \gamma_{nrad})/\gamma_{tot}$ (the black solid curve in Fig. 2d). Considering the extracted non-radiative decay rate, meta-mirrors A, B, and C modify the radiative decay rate, $\gamma_{rad} = F_P\gamma_{rad,0}$, from 0.97, 0.26, and 0.01 meV. Moreover, as demonstrated by its radiative and total decay rates, meta-mirror C suppresses the vacuum field to a near-zero value and causes the exciton lifetime to approach the non-radiative decay limit. We expect that the meta-mirror platform combined with high-quality TMDCs can support long-lived intralayer excitons for long-range interactions[3,4,5,6,7] and transport[9,10,11,12].

The meta-mirror brings its polarisation-dependent features to the radiative decay dynamics of excitons in the $MoSe_2$ monolayer. A meta-mirror consisting of anisotropic meta-atoms manipulates the vacuum field interference depending on the polarisation state (Fig. 3a). Upon employing the Au nanorods as meta-atoms, the exciton PL emission exhibits different radiative decay rates for the horizontal (H) and vertical (V) polarisation states. Fig. 3b shows the mechanism of polarisation-dependent vacuum field manipulation using the anisotropic meta-mirror. Depending on its length ($q$) and width ($p$), the anisotropic Au nanorod supports localised surface plasmon resonance (LSPR) at different frequencies for the H- and V-polarisation states. The higher the frequency of light, the longer will be the length/width for LSPR ($L_R$): $L_R$ of ~20 and ~100 nm for the pumping and PL emission frequencies (Supplementary Information S6). For the pumping frequency, both $p$ and $q$ are sufficiently larger than $L_R$, and the off-resonance behaviour causes a similar reflection amplitude and phase for the H- and V-polarised states. In contrast, for the emission frequency, the length (width) of the Au nanorod becomes larger (shorter) than $L_R$ by an amount $\Delta L$: $q = L_R+\Delta L$ and $p = L_R-\Delta L$. In this configuration, the Au nanorod reflects H- and V-polarised light with an equivalent reflectance but causes a significant difference in their reflection phases. Moreover, the relative phase difference can be easily tuned from the in-phase to out-of-phase conditions by changing the width and length. The high reflectance of the off-resonance state increases the visibility of the vacuum field interference.

We measured the birefringent reflectance of the fabricated anisotropic meta-mirror using polarisation-resolved off-axis holography (Fig. 3c). The meta-mirror consists of 71 × 35 Au nanorods with a period of 280 nm (See Method). At the pumping frequency, the reflection amplitude averaged over the meta-mirror was 0.82±0.051 and 0.79±0.050 for the H- and V-polarised states, respectively (Supplementary Information S7), and the reflection phase difference is only ~4.09°. The measured reflection amplitude and phase difference enable an approximately equivalent pumping condition independent of the polarisation state. At the emission frequency, the reflection phase difference between the H- and V-polarised states is ~128.06°, which shows nice consistency with the numerical calculation (Supplementary Information S6).

Subsequently, using the anisotropic meta-mirror, we demonstrated the polarisation-dependent customisation of the radiative decay of the neutral exciton in the $MoSe_2$ monolayer (Fig. 3d). The polarisation angle of the pumping light is 45°. The PL intensity of the V-polarised state is brighter than that of the H-polarised state inside the area of the meta-mirror. However, the PL intensity does not exhibit a notable difference outside the meta-mirror. We conducted a statistical analysis of the linewidth and intensity of the measured PL spectra depending on the polarisation state, and the results are illustrated in Fig. 3e. The measured linewidth is 2.70±0.21 and 2.53±0.26 meV for the V- and H-polarisation states, respectively. The ratio of the PL emission intensity of the V-polarisation to H-polarisation is ~1.30. Based on the relation between the PL intensity and exciton decay rate, we revealed that the radiative decay rate for the H- and V-polarisation states are 0.43 and 0.60 meV, respectively (Supplementary Information S8). We further experimentally confirmed that the same polarisation-dependent behaviour of the linewidth/decay rate appears when the polarisation of the pump light is set to be parallel to that of the emission light (Supplementary Information S9).

In this study, we demonstrated a meta-mirror platform to customise the radiative decay rate of TMDC excitons and its two-dimensional distribution by manipulating the reflection phase and amplitude of the vacuum field as desired. In a purely optical manner, we controlled the radiative decay rate of the neutral excitons of the monolayer MoSe$_2$ by two orders of magnitude, which resulted in a significant variation of the spectral linewidth from 1.28 to 2.24 meV. The seamless integration of the meta-mirror platform and TMDC material without affecting the intrinsic properties of the excitons enabled the experimental identification of the correlation between the emission intensity and spectral linewidth and the observation of the Lamb shift. The anisotropic meta-mirror assigned polarisation dependency to the radiative decay dynamics of the TMDC excitons. We expect that the position- and polarisation-dependent customisation of the radiative decay dynamics of excitons in two-dimensional space will be a promising approach for yielding exciton transport[9,10,11,12,34] and strong-correlation effects[3,4,5,6,7] with more optical degrees of freedom and demonstrate advanced photon–exciton transducers[35,36].

**Acknowledgement**

M.-K.S. acknowledges support from the KAIST Cross-Generation Collaborative Lab project and the National Research Foundation of Korea (NRF) (2020R1A2C2014685 and 2020R1A4A2002828). J.K. acknowledges the support of the National Research Foundation of Korea (NRF) (2020R1A2C2103166). J.S. acknowledges the support of the (NRF 2021R1A2C2008687). D.K. acknowledges support from the NRF (2015H1A2A1033753). Y.-S.C. acknowledges support from the NRF (2020R1I1A01069219). K.W. and T.T. acknowledge support from JSPS KAKENHI (Grant Numbers 19H05790, 20H00354 and 21H05233) and A3 Foresight by JSPS.

**Author Contributions**

S.P., D.K., and M.-K.S. conceived the project. S.P., D.K., and A.B. fabricated the meta-mirrors. S.P., D.K., D.K., and S.Y. fabricated the two-dimensional heterostructures. S.P. and Y.-S.C. performed the optical phase measurements. S.P. performed the confocal PL measurements. S.P. and D.K. performed the theoretical calculations. K.W. and T.T. provided the boron nitride crystals. S.P., D.K., and M.-K.S. analysed the data and wrote the manuscript with the assistance of all other authors.

## Method

### Sample Fabrication

To fabricate the meta-mirror platform for obtaining the results illustrated in Figs. 1 and 2, we initially deposited a 3-nm-thick Ti adhesive layer, a 200-nm-thick Au layer, a 60-nm-thick SiO$_2$ layer, and a 40-nm-thick Au layer onto the Si substrate sequentially via electron beam evaporation. To fabricate the meta-mirror platform for the polarisation-dependent exciton dynamics control, as represented in Fig. 3, we initially deposited a 100-nm-thick Al layer, a 40-nm-thick SiO$_2$ layer, and a 40-nm-thick Au layer onto the Si substrate sequentially using radio-frequency sputtering and thermal evaporation. The meta-mirrors were patterned using a polymethyl methacrylate (PMMA) layer via electron beam lithography. A 20-nm-thick Cr layer, deposited via electron beam evaporation, was used as a hard mask to transfer the meta-mirror patterns. Au disk arrays were fabricated via Ar ion-beam milling, and the Cr hard mask was removed using an etchant (CR-7). An HSQ (XR-1541, Dow Silicones Corp.) layer was spin-coated and thermally annealed, and the roughness of its top surface was ~2.72 nm in root-mean-square (Supplementary Information S3). Using adhesive tape, the hBN multilayers and MoSe$_2$ monolayer were physically exfoliated from the bulk crystals that were obtained from the National Institute for Materials Science (NIMS), Japan, and 2D Semiconductors Inc., respectively. The preparation and transfer of the van der Waals heterostructures were performed using the dry-transfer method employing polyethylene. Using an atomic force microscope (AFM), we measured the thickness of the bottom hBN layer, corresponding to the top-most layer of the meta-mirror, as 174 nm. The thickness of the hBN layer covering the MoSe$_2$ monolayer was measured as 6 nm.

### Numerical Simulation

We employed the finite-difference time-domain method (Lumerical, Ansys Canada Ltd.) for the numerical electromagnetic field calculations. The Drude-critical point model fits the experimentally determined complex refractive indices of Au[37], hBN[38], and HSQ[39]. The refractive index of the SiO$_2$ was fixed at 1.45. A periodic boundary condition was employed to calculate the reflection amplitude and phase pickup of the meta-mirrors. The 60-nm-thick SiO$_2$ layer was positioned onto the Au substrate. The period and radius of the Au nanodisks were measured from the obtained scanning electron microscope (SEM) images (Supplementary Information S2). The thicknesses of the HSQ and hBN layers were 140 nm and 180 nm, respectively. We measured the cross section and thickness of the HSQ layer using a SEM and the thickness of the hBN layer using an AFM.

**Optical measurement**

We employed a slightly off-axis holographic setup[40] based on an interferometric imaging system to measure the reflection amplitude and phase of the meta-mirrors. A light-emitting diode filtered by a narrow band-pass filter (530±5 nm for the pumping frequency and 760±5 nm for the PL emission frequency) illuminated the meta-mirrors through an objective lens with a numerical aperture (NA) of 0.40. We employed a confocal microscopy setup combined with a cryogenic chamber (CS204-DMX-20-OM, Advanced Research Systems, Inc.) for PL spectrum distribution measurements at 8 K. A two-axis galvanometer mirror system (GVS102, Thorlabs, Inc.) was used to scan the area of interest. A 532-nm-wavelength laser (OBIS, Coherent) was used to excite the PL emission and was blocked by a 700 nm long-pass filter (FELH0700. Thorlabs, Inc). A 20× long working-distance objective lens (NA = 0.40, Mitutoyo) collected the PL emission from the $MoSe_2$ monolayer to the spectrometer system consisting of a Czerny–Turner monochromator and a 2D array detector (SP2500 and PIXIS 100, Princeton Instruments). The used pumping power for the measurement of Fig. 2 was 17.7 µW (Supplementary Information S10).

**References (method)**

37. McPeak, K. M. et al. Plasmonic Films Can Easily Be Better: Rules and Recipes. *ACS Photonics*. 2, 326–333 (2015)

38. Lee, S.-Y., Jeong, T.-Y., Jung, S. & Yee, K.-J. Refractive Index Dispersion of Hexagonal Boron Nitride in the Visible and Near-Infrared. *Phys. Status Solidi B*. **256**, 1800417 (2019)

39. Yang, C.-C. & Chen, W.-C. The structures and properties of hydrogen silsesquioxane (HSQ) films produced by thermal curing. *J. Mater. Chem*. **12**, 1138–1141 (2002)

40. Guo, R. et al. LED-based digital holographic microscopy with slightly off-axis interferometry. *J. Opt.* **16**, 125408 (2014)

**References**

1. Mak, K.F., Xiao, D. & Shan, J. Light–valley interactions in 2D semiconductors. *Nat. Photon* **12**, 451–460 (2018)

2. Jia, Y. et al. Evidence for a monolayer excitonic insulator. *Nat. Phys.* **18**, 87–93 (2022)

3. Smoleński, T. et al. Signatures of Wigner crystal of electrons in a monolayer semiconductor. *Nature* **595**, 53–57 (2021).

4. Wang, Z. et al. Evidence of high-temperature exciton condensation in two-dimensional atomic double layers. *Nature* **574,** 76–80 (2019).

5. Arp, T.B. et al. Electron–hole liquid in a van der Waals heterostructure photocell at room temperature. *Nat. Photon.* **13**, 245–250 (2019)

6. Regan, E. C. et al. Optical detection of Mott and generalized Wigner crystal states in $WSe_2/WS_2$ moiré superlattices. *Nature*, 579,359–363 (2020)

7. Li, H. et al. Imaging two-dimensional generalized Wigner crystals. *Nature* **597**, 650–654 (2021).

8. Seyler, K. L. et al. Signatures of moiré-trapped valley excitons in $MoSe_2/WSe_2$ heterobilayers. *Nature* **567**, 66–70 (2019).

9. Onga, M., Zhang, Y., Ideue, T. & Iwasa, Y. et al. Exciton Hall effect in monolayer $MoS_2$. *Nature Mater* **16,** 1193–1197 (2017).

10. Unuchek, D. et al. Room-temperature electrical control of exciton flux in a van der Waals heterostructure. *Nature* 560, 340–344 (2018)

11. Kulig, M. et al. Exciton Diffusion and Halo Effects in Monolayer Semiconductors. *Phys. Rev. Lett*. **120**, 207401 (2018)

12. Unuchek, D. et al. Valley-polarized exciton currents in a van der Waals heterostructure. *Nat. Nanotechnol.* **14**, 1104–1109 (2019).

13. Lee, J., Mak, K. & Shan, J. Electrical control of the valley Hall effect in bilayer $MoS_2$ transistors. *Nature Nanotech* **11**, 421–425 (2016)

14. Lien, D.-H., et al. Large-area and bright pulsed electroluminescence in monolayer semiconductors. *Nat Commun* **9,** 1229 (2018)

15. Shreiner, R., Hao, K., Butcher, A., and High, A. A. Electrically controllable chirality in a nanophotonic interface with a two-dimensional semiconductor. *Nat. Photon.* **16**, 330–336 (2022).

16. Kim, D. and M.-K. Seo. Experimental Probing of Canonical Electromagnetic Spin Angular Momentum Distribution via Valley-Polarized Photoluminescence. *Phys. Rev. Lett*. **127**, 223601. (2021)

17. Niehues, I. et al. Strain Control of Exciton–Phonon Coupling in Atomically Thin Semiconductors. *Nano Lett*. **18**, 1751–1757 (2018)

18. Kumar, S., Kaczmarczyk, A. & Gerardot, B. D. Strain-Induced Spatial and Spectral Isolation of Quantum Emitters in Mono- and Bilayer $WSe_2$. *Nano Lett.* **15**, 7567–7573 (2015)

19. Amani, M. et al. Near-unity photoluminescence quantum yield in $MoS_2$, *Science*, **350**, 1065–1068 (2015)

20. Jauregui, L. A. et al. Electrical control of interlayer exciton dynamics in atomically thin

heterostructures. *Science*. **366**, 870–875 (2019)

21. Chernikov, A. et al. Electrical Tuning of Exciton Binding Energies in Monolayer $WS_2$. *Phys. Rev. Lett*. **115**, 126802 (2015)

22. Tang, H. et al. Electrically Controlled Wavelength-Tunable Photoluminescence from van der Waals Heterostructures. *ACS Appl. Mater. Interfaces*, **14**, 19869–19877 (2022)

23. Fang, H. H. et al. Control of the Exciton Radiative Lifetime in van der Waals Heterostructures. *Phys. Rev. Lett.* **123**, 067401 (2019)

24. Horng, J. et al. Engineering radiative coupling of excitons in 2D semiconductors. *Optica*. **6**, 1443-1448 (2019)

25. Zhou, Y. et al. Controlling Excitons in an Atomically Thin Membrane with a Mirror. *Phys. Rev. Lett.* **124**, 027401 (2020)

26. Rogers, C. et al. Coherent feedback control of two-dimensional excitons. *Phys. Rev. Research*. **2**, 012029 (2020)

27. Sun, Z., et al. Optical control of room-temperature valley polaritons. *Nat. Photon.* **11**, 491–496 (2017)

28. Dufferwiel, S. et al. Valley-addressable polaritons in atomically thin semiconductors. *Nat. Photon.* **11**, 497–501 (2017).

29. Chen, Y.-J., Cain, J., Stanev, T. K., Dravid, V. P. & Stern, N. P. Valley-polarized exciton–polaritons in a monolayer semiconductor. *Nat. Photon.* **11**, 431–435 (2017)

30. Meinzer, N., Barnes, W. & Hooper, I. Plasmonic meta-atoms and metasurfaces. *Nat. Photon.* **8,** 889–898 (2014)

31. Jun, Y., Huang, K. & Brongersma, M. Plasmonic beaming and active control over fluorescent emission. *Nat Commun* **2,** 283 (2011)

32. Zheng, G. et al. Metasurface holograms reaching 80% efficiency. *Nature Nanotech* **10,** 308–312 (2015)

33. Li, Z. et al. Tailoring $MoS_2$ Valley-Polarized Photoluminescence with Super Chiral Near-Field., *Adv. Mater.* **30**, 1801908 (2018)

34. Huang, Z. et al. Robust Room Temperature Valley Hall Effect of Interlayer Excitons. *Nano Lett.* **20,** 1345–1351 (2020)

35. Kim, H., Uddin, S. Z., Higashitarumizu, N., Rabani, E. & Javey, A. Inhibited nonradiative decay at all exciton densities in monolayer semiconductors. *Science*, **373**, 448-452 (2021)

36. Lien, D.-H. et al. Electrical suppression of all nonradiative recombination pathways in monolayer semiconductors, *Science*, **364**, 468- 471 (2019)

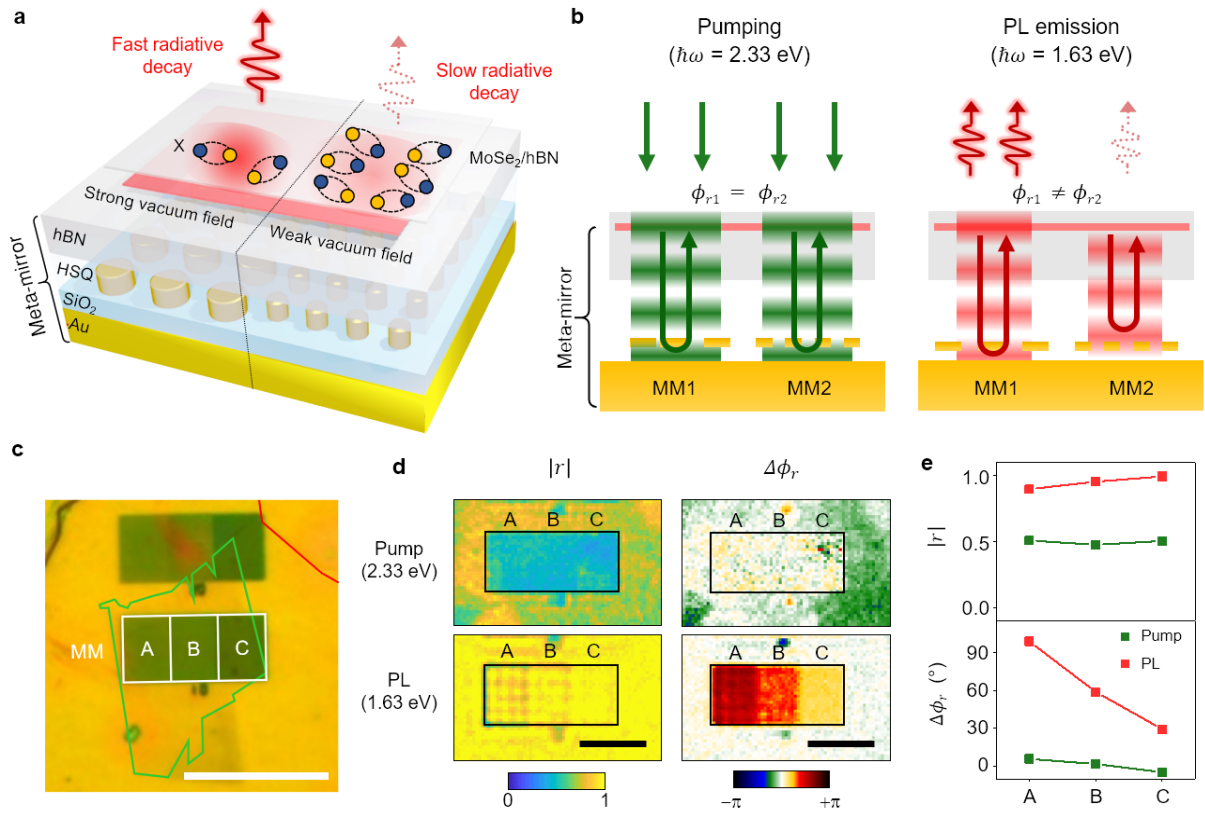

**Fig. 1 Engineering the vacuum-field interference and exciton decay dynamics by the meta-mirror platform. a, b,** Schematic illustration and mechanism of the radiative decay control of excitons in the meta-mirror platform. By controlling the vacuum field in the plane of the TMDC layer, the meta-mirror customises the two-dimensional landscape of the radiative decay and density of excitons. We present two representative meta-mirrors, MM1 and MM2, that support the same reflection phases ($\phi_{r1} = \phi_{r2}$) at the pumping frequency but different reflection phases ($\phi_{r1} \neq \phi_{r2}$) at the PL emission frequency. When the reflection amplitude is the same, the pumping condition is identical, but the Purcell effect on the radiative decay of PL emission is considerably different. **c,** Optical microscope image of the MoSe$_2$ monolayer (green solid polygon) on the meta-mirrors (white rectangles). The red solid line is a part of the boundary of the encapsulating hBN flake. Scale bar: 20 μm. **d,** Measured distribution of the reflection amplitude |r| and the relative reflection phase $\Delta\phi_r$ of the meta-mirrors at the pumping ($\hbar\omega$ = 2.33 eV) and emission (1.63 eV) frequency. Scale bar: 10 μm. **e,** Reflection amplitude and relative reflection phase averaged over the area of each meta-mirror A, B, and C for the pumping and PL emission frequencies.

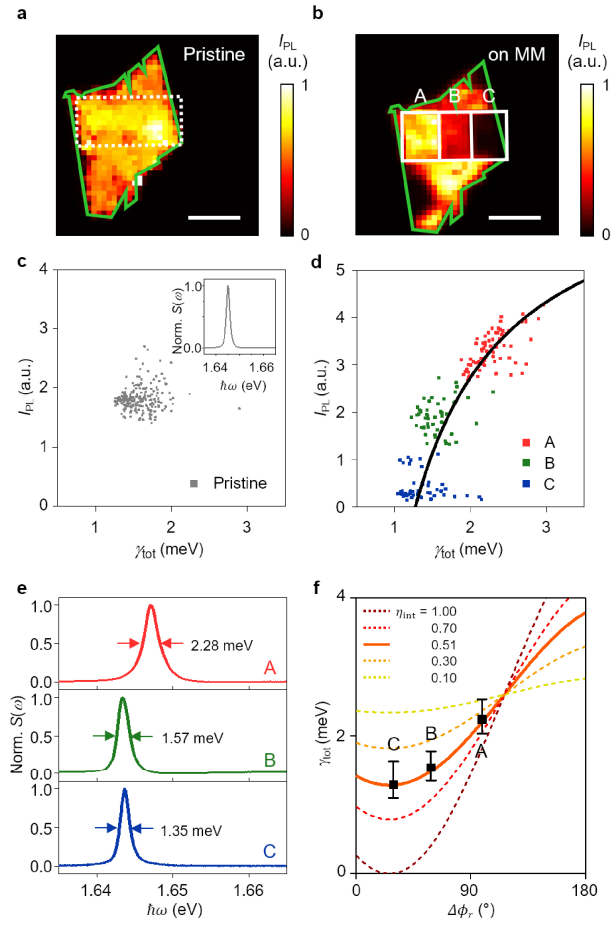

**Fig. 2 Manipulation of radiative decay dynamics of neutral excitons. a, b,** PL intensity distribution of neutral excitons of the hBN-encapsulated MoSe$_2$ monolayer on the Si/SiO$_2$ substrate (pristine) and the meta-mirrors. The green and white outlines indicate the region of the MoSe$_2$ monolayer and the meta-mirrors, respectively. Scale bar: 10 μm. **c, d,** Scatter plots of the intensity ($I_{PL}$) and linewidth ($\gamma_{tot}$) of the neutral exciton PL spectrum on **(c)** the Si/SiO$_2$ substrate and **(d)** meta-mirrors A, B, and C. The representative neutral exciton PL spectrum of the pristine MoSe$_2$ monolayer is shown in the inset of (c). $S(\omega)$ represents the normalised spectral density. **e,** The representative PL emission spectra of the neutral exciton engineered by the meta-mirror A, B, and C (the red, green, and blue solid lines, respectively). **f,** Results of the model analysis on the total decay rate ($\gamma_{tot}$) depending on the relative reflection phase shift ($\Delta\phi_r$). By employing the coherent optical feedback model, we calculated $\gamma_{tot}$ for different internal quantum efficiencies ($\eta_{int}$). The model with $\eta_{int}$ = 0.51 yields the best fit (the solid orange line) to the experimental measurements (the black squares and error bars).

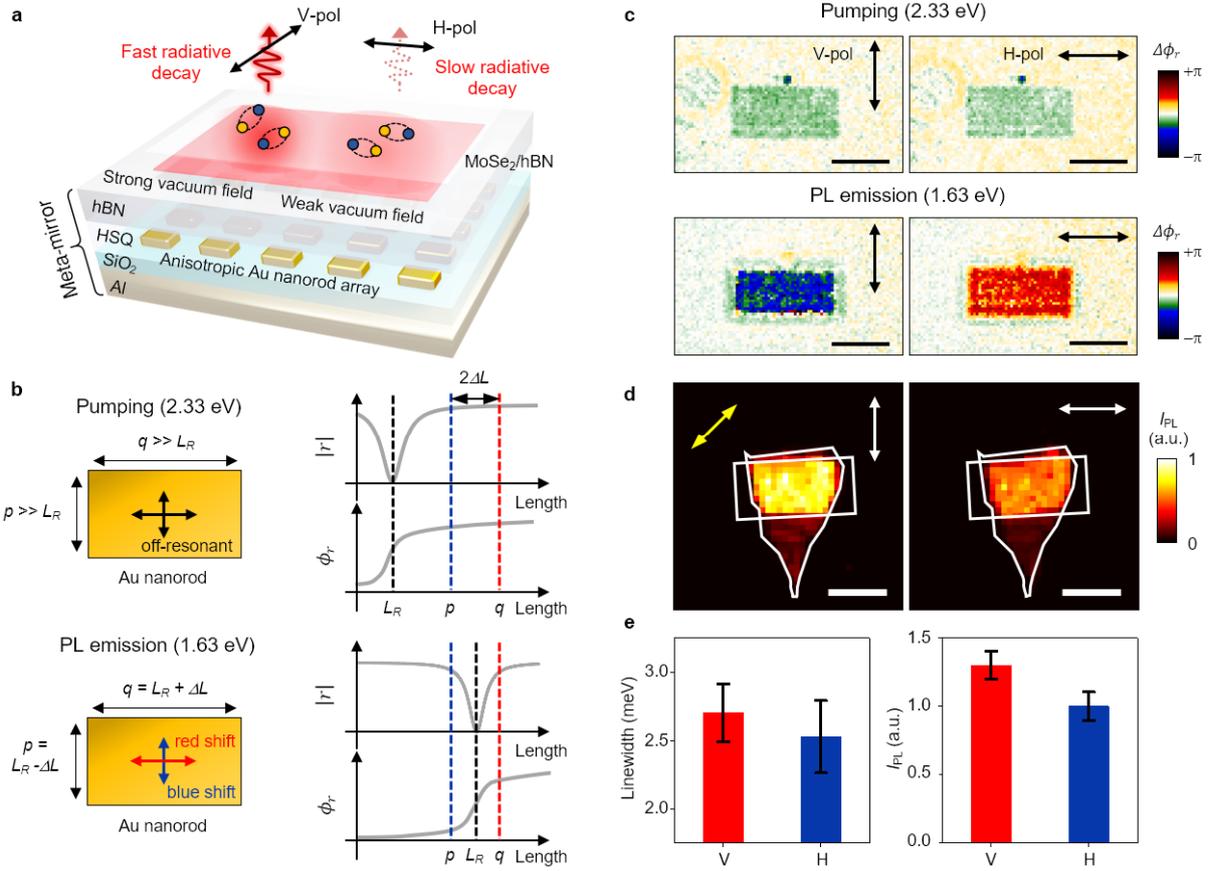

**Fig. 3 Polarisation-dependent control of exciton radiative decay dynamics. a,** Schematic illustration of the anisotropic meta-mirror for assigning the linear-polarisation dependency to the radiative decay dynamics of the TMDC excitons. **b,** Plasmonic response of the Au nanorod of the anisotropic meta-mirror. $L_R$ symbolises the length required for the Au nanorod to support the localised plasmon resonance. Depending on the choice of the length ($q$) and width ($p$) of the Au nanorod with respect to $L_R$, we can independently engineer the reflection phases for the H- and V-polarisation states. $\Delta L$ represents half of the difference between $p$ and $q$. **c,** The distribution of the relative reflection phase of the anisotropic meta-mirror at the pumping and emission frequency for the H- and V-polarisation. Scale bar: 10 μm. **d,** Measured PL emission intensity distribution of neutral excitons on the anisotropic meta-mirror for the V- (left panel) and H- (right panel) polarisation state. The yellow arrow indicates the polarisation of the pumping light. Scale bar: 10 μm. **e,** Measured PL spectrum linewidth (left panel) and intensity (right panel) of the neutral exciton emission depending on the polarisation state.